\documentclass[aps,twocolumn,prd,superscriptaddress,noshowpacs,nofootinbib,noshowkeys,floatfix]{revtex4}
\usepackage{graphicx}
\usepackage{amsfonts}
\usepackage{amssymb}
\usepackage{amsmath}

\usepackage{ulem}
\usepackage{color}

\renewcommand\sout{\bgroup \color{blue} \ULdepth=-.5ex \ULset}

\begin{document}
\title{Thermodynamics of Van der Waals fluids with quantum statistics}
\date{\today}
\author{Krzysztof Redlich}
\affiliation{Institute of Theoretical Physics, University of Wroc\l aw, PL-50-204 Wroc\l aw, Poland}
\affiliation{EMMI, GSI Helmholtzzentrum f\"ur Schwerionenforschung, 64291 Darmstadt, Germany}
\affiliation{%
Department of Physics, Duke University, Durham,  North Carolina 27708, USA}
\author{Kacper Zalewski}
\affiliation{Institute of Nuclear Physics, Polish Academy of Sciences, PL-31-342 Krak\'ow, Poland}
\affiliation{%
Institute of Physics, Jagiellonian University, PL-30-059 Krak\'ow, Poland}

\begin{abstract}
We consider thermodynamics of the van der Waals fluid  of quantum systems.
We derive  general  relations of
 thermodynamic functions and parameters of any ideal gas  and the corresponding van der Waals fluid. This provides  unambiguous  generalization of the classical van der Waals theory to quantum statistical systems.  As an example,   we apply the van  der Waals fluid  with fermi statistics to  characterize  the liquid-gas critical point in nuclear matter. We  also  introduce  the Bose-Einstein condensation in the  relativistic van  der Waals boson gas,  and argue, that it exhibits  two-phase  structure separated in space.
\end{abstract}
\maketitle

\section{Introduction}

{The well-known theory of the van der Waals fluid is a very useful extension of the {thermodynamics}  of classical ideal gases.
 It accounts for the repulsive interactions of particles with extended volume, as well as for the  inter-particle attractive force. In addition,
  it makes it possible to study phase transitions.  It is a  semi-quantitative theory  which has many  applications to model  the observed experimental behaviors of critical phenomena in the liquid-gas  phase  transition. In particular, under some approximations, it can be used as  a phenomenological tool to  describe    phase structure of  nuclear-,    or thermodynamical properties of  hadronic-matter \cite{k1,k2,r0,r1,GORI,GORII,BUG,GORIII,kap,heppe,kacper,VAO,VAG,PVA,VAGIII,VAGII}.

In general, the  van der Waals  equation of state  is restricted to describe   classical fluids, where the effects of quantum statistics are neglected.
  Recently,   an  interesting extension of the theory of quantum ideal gases has been proposed to account for repulsive and attractive interactions in the low density approximation \cite{VAG,PVA}. There, authors argued,   that a quantum generalization of the van der Waals equation of state is a non trivial task,  and proposed  three conditions  that such a generalization should satisfy.
{ The first two  are rather straightforward, as they require  that, (i)  in the limit where both repulsive and attractive interactions are negligible,   the new equation of state should coincide  with  that of  ideal quantum gases, and  (ii)  in the  limit of Boltzmann statistics,  it should  reproduce the classical van der Waals equation.} The third condition does not concern the equation of state {as it requires, that}  the entropy is  non-negative and tends to zero for $T\to 0$.  Based on the above conditions, the authors in Refs. \cite{VAG,PVA}, {have proposed an interesting approach   to formulate   the  theory of the van der Waals fluid with quantum statistics.}

In the present paper, by a slight extension of the original van der Waals argument we derive, unambiguously,  general relations between the thermodynamic functions and parameters of any ideal gas and those of its van der Waals generalization.
This allows for a  straightforward formulation of  thermodynamics and the  equation of state of the van der Waals fluids  with  quantum statistics.

 In the above   approach,  the conditions of Refs. \cite{VAG,PVA} are naturally satisfied.
  In particular, we show,  that {in quantum systems with the van  der Waals interactions,} the entropy does not change. Therefore, the entropy of the  van der Waals fluid tends to zero for $T\rightarrow 0$, if and only if,   the same is true for the corresponding ideal gas. Since {at constant volume and particle numbers,} the entropy is a non-decreasing function of temperature,    this also implies, that it is  non-negative.

 As an application we present  thermodynamic functions for the van der Waals fluids with quantum statistics,  and show  that for fermions,   our results agree with those obtained, previously in Ref. \cite{VAG}, by using different approach and methods. For the van der Waals fluid  of bosons we {argue},   that the Bose-Einstein condensation appears and  leads to two phases   which are separated in space.

\section{General analysis}

{Thermodynamic properties of a system  are characterized by an  equation of state which relates  different state variables associated with the matter. A familiar example is the  equation of state of the  non-relativistic  ideal gas  undergoing  classical or quantum statistics, }
\begin{equation}\label{eqs}
  p = p_{id}(V,T,N).
\end{equation}
{For  a classical thermodynamic ideal gas,  Eq. \ref{eqs} reduces to a familiar form, $p_{id}(V,T,N)V=NT$, where $N$ is the number of particles.}

The knowledge of the equation of state, however, is not {sufficient}  to find all
 {relevant}  thermodynamic functions and parameters, e.g.
{  the equations of state for the ideal gases of monoatomic and of diatomic molecules  are identical, but the temperature dependence of the  energy of the two gases differs.}

  The full {thermodynamic information} is   contained in the free energy {which is obtained by}  {integrating} the thermodynamic identity,
\begin{equation}\label{efnape}
\frac{\partial F(V,T,N)}{\partial V} = -p(V,T,N),
\end{equation}
 {as}
\begin{eqnarray}\label{fideal}
   F_{id}(V,T,N) & =& \int_V^{V_0}p_{id}(V',T,N)dV'\\ &+& f_{id}(V_0,T,{N}),\nonumber
 \end{eqnarray}
where $V_0$ is an  arbitrary  {volume parameter,} and $f_{id}$ is a function {which can}  not {be} determined by the equation of state.

{A very well known, and phenomenologically relevant,  extension of an ideal gas equation of state is the    van der Waals generalization,  which  accounts for the  repulsive and attractive interactions in a fluid.  Denoting    by $a$ and $b$, the parameters which  control  the strength of the attractive and repulsive interactions, respectively, the equation of state of  the van der Waals fluid, reads \cite{book} }

\begin{equation}\label{eqsw}
p(V,T,N)={{NT}\over {V-bN}}-an^2,
\end{equation}
where,  $n=N/V$, is the particle  density.

{{The  van der Waals equation of state  (\ref{eqsw}), can be obtained from the ideal gas  equation of state (\ref{eqs}),  by the following replacements}\footnote{It is assumed that there is no Bose-Einstein condensate}}

\begin{eqnarray}
 p_{id}(V-bN,T,N) = p(V,T,N) + an^2,\label{wjeden1} 
\end{eqnarray}
\begin{eqnarray}
 V_{id} = V-bN,\label{wjeden2}
\end{eqnarray}
where  $V$ is the volume of the van der Waals fluid, and $n(V,T,N)$,   the corresponding particle density. According to the  substitution   (\ref{wjeden2}),  the density in  an ideal gas, $n_{id}(V_{id},T,N)$ is related to  $n(V,T,N)$,  {as}

\begin{equation}\label{}
  \frac{1}{n_{id}(V-bN,T,N)} = \frac{1}{n(V,T,N)} - b.
\end{equation}
The replacements {in Eqs.} (\ref{wjeden1}) and (\ref{wjeden2}), applied in the ideal gas equation of state, $p_{id}(V_{id},T,N)V_{id}=NT$, yield indeed,    the standard  van der Waals equation of state (\ref{eqsw}).

{In the statistical systems,}
 knowing the equation of state  { is already sufficient to characterize the   phase transition. Indeed,}  supplementing it by the {conditions}
\begin{equation}\label{cricon}
 \frac{\partial p}{\partial n} = 0,\qquad \frac{\partial^2 p}{\partial n^2} = 0,
\end{equation}
{it is possible to  identify}  the critical point (CP), if any. {In the classical van der Waals fluid e.g., } one can express the  parameters, $n_c,T_c,p_c$,  {at the CP} by the {interaction} constants,   $a$ and $b$, as
\footnote{ In general, the coefficients can depend on the dimensionless parameter $\frac{a}{mb}$, where $m$ is the particle mass.}

\begin{equation}\label{}
  n_c = \frac{1}{3b},\qquad T_c = \frac{8a}{27b},\qquad p_c = \frac{a}{27b^2}.
\end{equation}
{From the equation of state, one can also calculate fluctuations of the  particle number, which are important characteristics of the phase transition \cite{VAGIII,fl1,fl2,fl3,fl4,fl5,fl6,fl7,fl8}. These are quantified by the corresponding  cumulants $\chi_k$ of order $k>0$, as}

\begin{equation}\label{}
 \frac{\chi_k}{ VT^{k-1}}
 =\frac{\partial^k p}{\partial \mu^k}|_T
= \left(n\frac{\partial}{\partial p}\right)^k p|{_T},
\end{equation}
{where to get the second equality, we have used}  the Gibbs-Duhem equation.

In order to find, however,  all  thermodynamic functions of the van der Waals gas, {one needs} the free energy. Integrating the equation of state of the van der Waals fluid as in Eq (\ref{efnape}), {one finds}

\begin{eqnarray}\label{eqi1}
  F(V,T,N) & =& \int_V^{V_0}\left(p_{id}(V'-bN,T,N)- a\frac{N^2}{V'^2}\right)dV'\nonumber \\&+& f(V_0,T,N),
\end{eqnarray}
where $f$ is {an integration "constant".}

 On the other hand, the  Eq.  (\ref{fideal}) for the ideal gas,  can be rewritten, {as}

\begin{eqnarray}\label{eqi}
  F_{id}(V-bN,T,N) &=& \int_V^{V_0+bN}p_{id}(V'-bN,T,N)dV'\nonumber \\&+&f_{id}(V_0,T,N),
\end{eqnarray}
{In  the following, we} assume,  that
{ for sufficiently large  $V_0$,}
the upper limit of the integration in Eq. (\ref{eqi})  can be replaced by $V_0$. A sufficient condition is  that, in the low density limit, $p_{id}(V,T,N)$  is proportional to $\frac{1}{V}$. The two preceding {equations,}
 can  be combined,  to give

\begin{eqnarray}\label{}
&&F(V,T,N) - F_{id}(V-Nb,T,N) =  \nonumber\\ &=& \int_V^\infty\left[p(V',T,N) - p_{id}(V'-bN,T,N)\right]dV'\nonumber
   \\   &+& f(\infty,T,N)   - f_{id}(\infty,T,N)\nonumber \\ &=& -a\frac{N^2}{V} + \left[f(\infty,T,N) - f_{id}(\infty,T,N)\right].
\end{eqnarray}
 {In the spirit of the van der Waals approach, we}  make the assumption
\begin{equation}\label{warune}
  f(\infty,T,N) = f_{id}(\infty,T,N).
\end{equation}
{Consequently,}

\begin{equation}\label{fidvvw}
  F(V,T,N) = F_{id}(V-bN,T,N) - aNn.
\end{equation}
The above equation relates the free energy of the ideal and  van der Waals gas, and thus allows to extract all relevant thermodynamic quantities in a transparent way.
In particular,
{d}ifferentiating relation (\ref{fidvvw}) with respect to the temperature,  {one gets the relation between entropies}
\begin{equation}\label{relent}
  S(V,T,N) = S_{id}(V-bN,T,N).
\end{equation}
It is {rather transparent,}  that conditions (\ref{warune}) and (\ref{relent}) are equivalent. {Indeed, }in the van der Waals picture, the effect of the excluded volumes around the particles is assumed to be compensated by a suitable increase of the total volume. If so, the number of states accessible to the gas, and consequently {its}  entropy, should remain unchanged. Another argument is that,  since the replacements  (\ref{wjeden1}) and (\ref{wjeden2}) leave the temperature and the number of particles unchanged, there is no reason to expect, {that}  $f(T,\infty) \neq f_{id}(T,\infty)$.

{Following}  Eqs.   (\ref{relent}), and  {applying}   the thermodynamic identity,
$
  E = F +TS
$,
 {one gets the  energy of the van der Waals fluid,}
\begin{equation}\label{relene}
 E(V,T,N) = E_{id}(V-bN,T,N)-a\frac{N^2}{V}.
\end{equation}
Finally, differentiating both sides of Eq.  (\ref{fidvvw}) with respect to $N$, {we obtain the  following  relation between  chemical potentials of the ideal and the van der Waals gases,}
\begin{eqnarray}\label{relamu}
 \mu(V,T,N) &=& \mu_{id}(V-bN,T,N)     \\&+& bp_{id}(V-bN,T,N) - 2an(V,T,N).\nonumber
\end{eqnarray}

Let us {notice} that the van der Waals equation of state  can be justified from statistical physics  only in the low density limit. It is an intuitive extrapolation, though a very useful one.

\section{Van der Waals fluids with quantum statistics}

{It the previous   section, we have established the relation between the free energy, and some other state variables,   of the ideal   and the van der Waals gas.
Following previous  discussions  in  Refs. \cite{VAG} and  \cite{PVA}, we focus now,   on the quantum statistics generalization of the van der Waals fluid. We assume  first, that for bosons,  there is no  Bose-Einstein condensate contribution.}

Let us consider the ideal gases with quantum statistics.
{In general,} the equation of state for quantum ideal gases is given in a parametric form:

\begin{equation}\label{qsdens}
  \frac{N}{V} = \frac{d}{2\pi^2}\int_0^\infty \frac{k^2dk}{e^{\beta[\epsilon(k)-\mu_{id}]}+\eta},
\end{equation}

\begin{equation}\label{qspres}
p(V,T,N) = \frac{d}{6\pi^2}\int_0^\infty \frac{k^4}{e^{\beta[\epsilon(k)-\mu_{id}]}+\eta}\frac{dk}{\epsilon(k)},
\end{equation}
where
$
  \beta = {1}/{T},$  $\epsilon(k) = \sqrt{k^2+m^2}$  is the particle energy, and
$d$ its degeneracy factor,  while $\eta = 1$ for fermions and $\eta= -1$ for bosons.

 For a study of this  equation of state, {the chemical potential is a derived variable.}
  For given values of  $V,T,N$, the parameter $\mu_{id}$ can be calculated from equation (\ref{qsdens}), and substituting it into  Eq. (\ref{qspres}),  the corresponding value of $p(T,V,N)$ {can be}  found.

For the van der Waals fluids with quantum statistics,  we {derive}  the equations of state by making the substitutions (\ref{wjeden1}) and (\ref{wjeden2}) in Eqs. (\ref{qsdens}) and (\ref{qspres}).   { Consequently, we  obtain,    the following parametric equation of state:}

 \begin{equation}\label{qvwden}
  \frac{N}{V-bN} = \frac{d}{2\pi^2}\int_0^\infty \frac{k^2dk}{e^{\beta[\epsilon(k)-\overline{\mu}]}+\eta}
\end{equation}
and

\begin{equation}\label{qvwpre}
p(V,T,N) = \frac{d}{6\pi^2}\int_0^\infty \frac{k^4}{e^{\beta[\epsilon(k)-\overline{\mu}]}+\eta}\frac{dk}{\epsilon(k)} - a\frac{N^2}{V^2},
\end{equation}

The parameter $\mu_{id}$ in Eq. (\ref{qsdens}) is the chemical potential of the ideal gas. Therefore, comparing with equation (\ref{qvwden}),  {one finds, that}

\begin{equation}\label{vdwpot}
  \overline{\mu} = \mu_{id}(V-bN,T,N),
\end{equation}
 is not the chemical potential (\ref{relamu}) of the quantum van der Waals fluid.

 Equations (\ref{relene}) and (\ref{relamu}) can be used to calculate the energy and the chemical potential of the quantum van der Waals fluids in terms of the thermodynamic functions and parameters of the corresponding ideal gases. {
Furthermore, according to Eq.  (\ref{relent}),  the entropy of the quantum van der Waals fluids   vanishes at $T=0$.}

{To illustrate  the importance of the  effects of quantum statistics on the van der Waals equation of state and its critical properties, we consider its application to the description of the liquid-gas phase transition in nuclear {matter} \cite{k1,k2,lg1,lg2,lg3}.
Following Ref.  \cite{VAG} we model the  nuclear  matter as a gas  of nucleons.  Thus,  in Eqs. (\ref{qvwden}) and  (\ref{qvwpre}) the parameters are chosen as: $ \eta = +1, d=4$ and  $m = 0.939$ {GeV}.
In addition,  to fix the interaction strength parameters,  we take  as  inputs,
at $T=0$ and  $p=0$,  the experimental values of the particle density $n_0 = 0.16$ {fm}$^{-3}$, and the  energy per nucleon ${\epsilon}_0 = \frac{E}{N} = 0.922$ {GeV}.

The issue  is,  to find the density $n_c$, temperature $T_c$ and pressure $p_c$   at the critical point,  in  the  van der Waals fluid with quantum statistics.}

{We first determine}  the parameters $a$ and $b$  from the input data.
{ To proceed, it is convenient to define  the following functions:}

\begin{eqnarray}
  f_n
   &=& \frac{d(-1)^nn!}{6\pi^2}\int_0^\infty\frac{dk k^4}{\epsilon(k)}  \frac{e^{\beta n(\epsilon(k)-\overline{\mu})}}{(e^{\beta (\epsilon(k)-\overline{\mu})} + \eta)^{n+1}} \\
  g_n
  &=& \frac{d(-1)^nn!}{2\pi^2}\int_0^\infty\frac{dk k^2 e^{\beta n(\epsilon(k)-\overline{\mu})}}{(e^{\beta (\epsilon(k)-\overline{\mu})} + \eta)^{n+1}} \\
  \epsilon_{id}
   &=& \frac{d}{2\pi^2}\int_0^\infty\frac{\epsilon(k)k^2dk}{e^{\beta (\epsilon(k)-\overline{\mu})} + \eta}.
\end{eqnarray}
{Then, }
Eqs.   (\ref{qvwden}), (\ref{qvwpre}) and (\ref{relene}), {applied at}   $T=0$ and $p=0$, {yield}:
\begin{eqnarray}
  \frac{n_0}{1-bn_0} &=& g_0(0,\overline{\mu}), \\
  p &=& f_0(0,\overline{\mu}) - an_0^2 = 0, \\
  \overline{\epsilon}_0 &=& \epsilon_{id}(0,\overline{\mu})(n_0^{-1} - b) - an_0.
\end{eqnarray}
 {Thus,  the value of  $\overline{\mu}$,  is obtained from:}

\begin{equation}\label{}
  n_0(\epsilon_{id}(0,\overline{\mu})-g_0(0,\overline{\mu})\overline{\epsilon}_0) = f_0(0,\overline{\mu})g_0(0,\overline{\mu}).
\end{equation}
Its solution, $\overline{\mu} = 0.9994$GeV, substituted into the first two equations, yields

\begin{equation}\label{}
a = 0.3291\mbox{GeVfm}^3, \qquad b = 3.416\mbox{fm}^3,
\end{equation}
in agreement with the results of ref. \cite{VAG}.

In order to find the  parameters  {at the CP} we need, {in addition to}  the equation of state (\ref{qvwden}) and (\ref{qvwpre}), also  {relations} (\ref{cricon}). For the present case,  they can be written in {the following} form

\begin{eqnarray}
  f_1(T,\overline{\mu})(1+bg_0(T,\overline{\mu}))^3 - 2ag_0(T,\overline{\mu})g_1(T,\overline{\mu}) =0, 
\end{eqnarray}
\begin{eqnarray}
    \frac{f_2(T,\overline{\mu})}{f_1(T,\overline{\mu})} - \frac{g_2(T,\overline{\mu})}{g_1(T,\overline{\mu})} + \frac{g_1(T,\overline{\mu})}{g_0(T,\overline{\mu})}\frac{2bg_0(T,\overline{\mu})-1}{1+bg_0(T,\overline{\mu})} = 0.
\end{eqnarray}
Solving this pair of equations for $T$ and $\overline{\mu}$,  {yields}

\begin{equation}\label{r1}
  T_c = 19,72\mbox{MeV},\qquad \overline{\mu}_c = 0.9484\mbox{GeV}.
\end{equation}
Substituting these  {values of the critical temperature and the parameter $\overline{\mu}_c$ } into the equation of state,  we get

\begin{equation}\label{r2}
  n_c = 0.072\mbox{fm}^{-3},\qquad p_c = 0.526\mbox{MeVfm}^{-3},
\end{equation}
{as the critical density and pressure at the CP  of  the nuclear matter,  modeled as the van der Waals fluid with quantum statistic.}

 The results  in Eqs. (\ref{r1}) and (\ref{r2})  agree   with those  obtained in Ref. \cite{VAG}, showing  that our  formulation of the van  der Waals model with  quantum statistics is quantitatively  consistent with  previous findings.

It is worth noting,  that in the van der Waals model  with Boltzmann
statistics,  and at the same values of $a$ and $b$, the critical temperature  at the CP is by almost ten degrees higher. Thus, in the application of  this model   to  the  liquid-gas phase transition in nuclear physics, the quantum statistic effects are important,  and make  the location of  the CP is closer  to that extracted  in the experiments.

\section{Bose-Einstein condensation  in the van der Waals  quantum fluids}

The van der Waals model is not sufficiently detailed to describe quantitatively the Bose-Einstein condensation,   a qualitative description, {however,} can be given.

Let us  {consider first}  the ideal gas of bosons. As {it is} seen from  Eq.  (\ref{qsdens}), the particle density, at {any} given temperature,  increases  {with}  the chemical potential  up to $\mu_{id}^c = m$, and beyond that value   the integral becomes divergent. {Consequently, when}  more particles are introduced into the system,  they form a condensate, i.e. a fluid of particles distributed uniformly over the volume $V$, where each particle is in its ground state, i.e. has momentum zero.

 Let us denote by $n_c$ the density of particles from the condensate,  and by $n_a,$  the  density of  {\it  active} particles with the momentum distribution given by the integrand in Eq. (\ref{qsdens}). Thus,

\begin{equation}\label{}
  n = n_a + n_c,
\end{equation}
{is the total density of particles in the system.}

The particles from the condensate do not strike the walls,  and therefore,  do not contribute to the pressure. {Consequently,}   {after}  $n_a$ reaches its maximum value,  the pressure does not depend on the particle density $n$. Therefore, the gas satisfies the thermodynamic stability condition

\begin{equation}\label{stacon}
 \left( \frac{\partial p}{\partial n}\right)_T \geq 0.
\end{equation}

For the van der Waals gas with a condensate,  the  substitution in Eq. (\ref{wjeden2}) remains valid, whereas Eq. (\ref{wjeden1})  becomes

\begin{equation}\label{}
  p_{id}(V-bN,T,N) = p(V,T,N)  + ann_a.
\end{equation}
{This is} because,  only the active particles hit the walls, while all the particles contribute to the force pulling back a particle about to hit the wall. In the region considered, the pressure $p_{id}$,  does not depend on the total particle density. Consequently, the pressure $p$,  decreases with increasing $n$. This contradicts the stability condition (\ref{stacon}).

 In order to see how the system collapses, let us  consider a vessel, containing the fluid with the condensate, divided into two parts $A$ and $B$ by a mobile wall impermeable to particles. We start with the fluid with both $n_a$ and $n_c$ {being} constant. Suppose now,  that the wall has moved by fluctuations, so that, volume $A$ increases a little at the expense of volume $B$. Consequently, in $A$,  the number of places for active particles has increased and the amount of condensate has decreased. The density of active particles does not change, but the density $n$ gets reduced because the volume has increased. As a result,  the
pressure of the fluid on the walls in $A$ increases. A similar argument shows, that the pressure in $B$,  decreases. Thus,  $A$ goes on growing and $B$ shrinking. The particles of the van der Waals fluid, however, have finite volumes. Therefore, when the particle density in $B$ becomes of the order of ${1}/{b}$, an additional
contribution to the pressure in $B$ appears,  and the process gets stopped. Another wall can be introduced into the increased volume $A$,  and the process may be repeated. Let us keep the notation $A$,  for the expanding part of the volume. After enough repetition, the  region $A$ will contain only active particles. This is an implication of the fact,  that for the van der Waals fluid, as opposed to the ideal gas, a homogenous phase consisting of both active  and condensate particles is unstable.
{Thus, in the van der Waals fluid with Bose-Einstain's condensate, one expects the appearance of the   two separated phases.  }

The above discussion,   is not enough to conclude,  whether {the}  two,  well-separated phases will be formed in the system,  or one of the phases  will  be dispersed in the form of small droplets, in the other. The need for the  separation, however, follows directly from the relation between the pressure and the total particle density.

\section{Conclusions}

The  objective of these studies was to establish  the generalization of the van der Waals fluids to  account for the quantum statistics.
The main result  is  { the transparent and unambiguous }  derivation of {the} relation {between} the free energy of any ideal gas  {and}  the corresponding van der Waals fluid.

 In order to establish such   generalization, it is not {enough}  to use the equations of state, but
{some} additional assumptions are needed.
The reason is rather straightforward,  since  the free energy, expressed as a function of thermal  variables, defines all {relevant}  thermodynamic functions of the system, but the equation of state does not.
{The additional assumption  needed, was  that  in the large volume limit,  the free energy of the ideal and van der Waals gas coincide.}

 The theory based on the relation between  the free energy of the ideal gases and van der Waals fluid provide complete description of the equation of state and thermodynamics of quantum systems.
For practical applications  it also implies  transparent  relations between  entropies,   energies  and  chemical potentials of quantum ideal gases and the van der Waals fluid.

The simplicity of our approach is illustrated by discussing the van der Waals   gas with Fermi  statistics and its application to a phenomenological description of liquid-gas  critical point in the nuclear matter. The Bose-Einstein condensation in the van der Waals gas, was argued to  imply 
 the structure of two phases separated
  in space. This is because, in the presence of the condensate, the stability condition which  requires non-decreasing  pressure with density at  constant temperature, which is fulfilled in  the ideal Bose gas, is not satisfied by the van der Waals quantum system.

The quantum effective theory of fluids proposed here,   was shown to satisfy all necessary conditions introduced in ref. \cite{VAG}  to  generalize    the van der Waals thermodynamic system to quantum statistics, however,
 these conditions are not enough to formulate it. }

\section*{ Acknowledgments}
One of the authors (KZ) was partly supported by the Polish National Science Center (NCN), under  grant DEC-2013/09/B/ST2/00497. K.R. acknowledges  support of  the Polish Science Center (NCN) under Maestro grant DEC-2013/10/A/ST2/00106,
and of the U.S. Department of Energy under Grant No.  DE-FG02- 05ER41367.

\end{document}